\documentclass[12pt,epsfig,epsf]{article}
\usepackage{amsmath}
\usepackage{cite}
\usepackage{slashed}
\usepackage{epsf,color,colordvi}
\usepackage{caption}
\usepackage[table]{xcolor}     
\usepackage{array} 
\usepackage{xspace}
\usepackage{graphicx}          
\usepackage{colortbl}          
\usepackage{booktabs}          
\usepackage{amssymb}
\usepackage{tabularx}
\usepackage{enumerate}
\usepackage{epsfig}
\usepackage{cite}
\usepackage{lscape}
\usepackage{makecell}
\usepackage{fontenc}
\usepackage{float}
\usepackage[lofdepth,lotdepth,caption=false]{subfig}
\usepackage{xcolor}
\usepackage{multirow}
\usepackage{longtable}
\usepackage{adjustbox}
\usepackage{indentfirst}
\usepackage{dcolumn}
\usepackage{rotating}
\usepackage[english]{babel}
\usepackage[autostyle]{csquotes}
\usepackage{hyperref}
\usepackage[normalem]{ulem}
\usepackage{cancel} 
\usepackage{color}
\graphicspath{{Figures/}}
\newcommand {\ignore}[1]{}
\definecolor{darkgreen}{cmyk}{1,0,1,0.4}
\definecolor{brown}{cmyk}{0,0.8,1,0.2}
\definecolor{darkred}{cmyk}{0,1,1,0.2}

\newcommand{\bi}{\begin{itemize}}
\newcommand{\ei}{\end{itemize}}

\def\beq{\begin{equation}}
\def\eeq{\end{equation}}

\newcommand{\bea}{\begin{eqnarray}}
\newcommand{\eea}{\end{eqnarray}}

\newcommand{\ie}{{\it i.e.}}

\def\epsilon{\varepsilon}

\newcommand\sch{Schr$\ddot{\rm o}$dinger~}

\def\ket#1{| \,#1\, \rangle}
\def\bra#1{\langle \,#1\, |}
\def\scx#1#2{\langle \,#1\, |\, #2\, \rangle}

\def\me#1#2#3{\langle \,#2\, |\,#1\,|\, #3\,\rangle}

\def\<{\langle}
\def\>{\rangle}

\def\dfrac#1#2{{\displaystyle\frac{#1}{#2}}}

\def\lsim{\mathrel{\rlap{\lower4pt\hbox{\hskip1pt$\sim$}}
    \raise1pt\hbox{$<$}}}         
\def\gsim{\mathrel{\rlap{\lower4pt\hbox{\hskip1pt$\sim$}}
    \raise1pt\hbox{$>$}}}         

\newcommand{\Dm}{\Delta m^{2}}
\newcommand{\mbar}{\overline{m}^{2}}

\newcommand{\dacp}[1]{\ensuremath{\delta [\Delta P^{CP/T}_{\alpha\beta}]}}

\newcommand{\pbarab}[1]{\ensuremath{{ P}_{\bar{\alpha} \bar{\beta}} }}
\newcommand{\pbarba}[1]{\ensuremath{{ P}_{\bar{\beta} \bar{\alpha}} }}
\newcommand{\acpab}[1]{\ensuremath{A^{CP}_{\alpha \beta}}}
\newcommand{\acpaa}[1]{\ensuremath{A^{CP}_{\alpha \alpha}}}
\newcommand{\ataa}[1]{\ensuremath{A^{T}_{\alpha \alpha}}}
\newcommand{\acpba}[1]{\ensuremath{A^{CP}_{\beta \alpha}}}
\newcommand{\atab}[1]{\ensuremath{{A}^{T}_{\alpha \beta}}}
\newcommand{\atba}[1]{\ensuremath{{A}^{T}_{\beta \alpha}}}
\newcommand{\acptab}[1]{\ensuremath{A^{CPT}_{\alpha \beta}}}

\newcommand{\dpcpme}[1]{\ensuremath{\Delta { P}^{CP}_{\mu e} }}
\newcommand{\dpcpmt}[1]{\ensuremath{\Delta { P}^{CP}_{\mu \tau} }}
\newcommand{\dpcpet}[1]{\ensuremath{\Delta { P}^{CP}_{e \tau } }}
\newcommand{\dpcpee}[1]{\ensuremath{\Delta { P}^{CP}_{e e} }}
\newcommand{\dpcpmm}[1]{\ensuremath{\Delta { P}^{CP}_{\mu \mu} }}
\newcommand{\dpcptt}[1]{\ensuremath{\Delta { P}^{CP}_{\tau \tau } }}
\newcommand{\dptme}[1]{\ensuremath{\Delta { P}^{T}_{\mu e}}}
\newcommand{\dptmt}[1]{\ensuremath{\Delta { P}^{T}_{\mu \tau}}}
\newcommand{\dptet}[1]{\ensuremath{\Delta { P}^{T}_{e \tau}}}
\newcommand{\dpcptab}[1]{\ensuremath{\Delta { P}^{CPT}_{\alpha \beta} }}
\DeclareMathOperator{\sech}{sech}

\hypersetup{
  colorlinks,
  citecolor=red,
  linkcolor=blue,
  urlcolor=blue}

\begin{document}
\begin{titlepage}

\vspace*{-3.cm}
\begin{flushright}

\end{flushright}


\renewcommand{\thefootnote}{\fnsymbol{footnote}}
\setcounter{footnote}{0}

{\begin{center}
{\large\bf 
 Two flavor neutrino oscillations in presence of 
 non-Hermitian dynamics  
\\[0.2cm]
}
\end{center}}

\renewcommand{\thefootnote}{\alph{footnote}}

\vspace*{.8cm}
\vspace*{.3cm}
{
\begin{center} 

   {\sf                 Kritika Rushiya$^{\S}$\, , 
                }
        {\sf               Gaurav Hajong$^{\dagger}$\,  ,
        }
           {\sf               Bhabani Prasad Mandal$^{\dagger}$\,    } and              
            {\sf                 Poonam Mehta$^{\S}$\,
            
}
\end{center}
}
\vspace*{0cm}
{\it 
\begin{center}

$^\S$\, School of Physical Sciences, Jawaharlal Nehru University, 
    New Delhi 110067, India  \\
    
$^\dagger$\, Department of Physics, Banaras Hindu University, Varanasi, 221005, India  \\

\end{center}
}

\abstract{We develop a consistent mathematical framework for studying two flavor neutrino oscillations in presence of non-Hermitian dynamics. We consider two approaches : (a) bi-orthonormal inner product defined by a positive-definite metric operator $\mathcal{G}$ and (b) the density matrix prescription by Brody and Graefe~\cite{Brody:2012nxf}. For the $\mathcal{PT}$-symmetric case, we show that the $\mathcal{G}$ metric approach does not work well  (probabilities are not conserved) both in $\mathcal{PT}$-unbroken as well as $\mathcal{PT}$-broken regime. 
Hence, we adopt the density matrix prescription by Brody and Graefe which is a positive semi-definite map. 
In the density matrix prescription, we note that probability in the steady state limit is not necessarily $1/2$ thereby indicating non-Markovian behavior. }

\vspace*{.5cm}

\end{titlepage}
\section{Introduction}
\label{sec:introduction}
Bender, in their seminal work \cite{Bender:1998ke}(see also~\cite{Bender2004,Bender:2005hf,Bender:2007nj,Bender:2007gg,Bender:2015aja,Bender:2017hqr,Bender:2019cwm,Bender:2021fxa,Bender:2023cem,Bender:2024eoq,duarte2017quantum}) introduced a new class of $\mathcal{PT}$-symmetric Hamiltonians which are invariant under the combined action of parity ($\mathcal{P}$) and time-reversal ($\mathcal{T}$) transformations. For any $\mathcal{PT}$-symmetric system, we can identify two distinct regimes, namely, $\mathcal{PT}$-unbroken regime and $\mathcal{PT}$-broken regime. When the eigenvalues along with the eigenvectors coalesce, we have the exceptional point  marking the  transition between the two regimes~\cite{Heiss_2000,Berry:2004ypy,Heiss:2012dx,Guenther:2007qkm,Brody:2013lbg}. In the $\mathcal{PT}$-unbroken regime, the eigenvalues are real while in the $\mathcal{PT}$-broken regime, the entire spectrum  or a part  of it become complex and complex eigenvalues appear in complex conjugate pair. The eigenvectors no longer form an orthonormal basis under the conventional Dirac inner product. We need to define an appropriate inner product that restores orthonormality, preserves the probability and maintains the unitarity of the theory. In literature, $\mathcal{PT}$ inner product and the $\mathcal{CPT}$  {inner product}\footnote{$\scx{\psi}{\phi}_{\mathcal{CPT}}$ = $(\mathcal{CPT}\ket{\psi})^{T} \cdot \ket{\phi}$~\cite{Ohlsson:2019noy}} have been commonly adopted in the discussion of $\mathcal{PT}$-symmetric Hamiltonians (see, for example,  ~\cite{Ohlsson:2015xsa,Ohlsson:2019noy,Ohlsson:2020gxx,Ohlsson:2020idi,Alexandre:2023afi}). 
It turns out that the $\mathcal{PT}$ inner product is not always positive-definite~\cite{Bender2004}. 
It was shown in~\cite{Ohlsson:2019noy,Ohlsson:2020idi}, that the framework using $\mathcal{CPT}$ {inner product}  works well  in the $\mathcal{PT}$-unbroken regime  leading  to consistent quantum theory for non-Hermitian systems. It may be noted that this claim relied on using the $\mathcal{CPT}$ eigenstates, which have no physical basis. However, in the $\mathcal{PT}$-broken regime,  the authors found inconsistencies  leading to non-conservation of probabilities~\cite{Ohlsson:2019noy,Ohlsson:2020idi}. This calls for devising proper treatment which provides a consistent framework for non-Hermitian Hamiltonians. In the present work we shall adopt two approaches to compute neutrino oscillation probabilities in presence of non-Hermitian dynamics, {\it{viz.}},
\begin{itemize}
\item  Bi-orthonormal inner product defined by positive-definite metric operator $\mathcal{G}$~\cite{Ju:2019kso,Tzeng:2020mwv,Mannheim:2017apd,Shi:2009pc,Kleefeld:2009vd,Shukla:2022kda} and
\item   Density matrix prescription by Brody and Graefe~\cite{Brody:2012nxf} (see also~\cite{Varma:2020tqu}).
\end{itemize}

$\mathcal{PT}$-symmetric systems  span diverse areas such as quantum information theory, quantum optics, condensed matter physics and high energy physics. In Ref.~\cite{Kawabata:2018edo}, it was shown that the  interplay between topological superconductivity and $\mathcal{PT}$-symmetry leads to anomalous particle transport. $\mathcal{PT}$-phase transitions have been studied in the context of anisotropic simple harmonic oscillator in 2d and charged isotropic oscillator in 3d under the influence of external imaginary magnetic field~\cite{Mandal:2013hd}. 
 {{A  formulation of non-Hermitian dynamics in terms of Bloch equation for a spin $1/2$ system, along with its implications for extreme violations of the quantum speed limit, was developed in~\cite{Varma:2020tqu}. This analysis, carried out in the context of $\mathcal{PT}$-symmetric dynamics, demonstrated how non-Hermiticity can dramatically accelerate quantum evolution. This was later extended to the $\mathcal{PT}$-broken regime~\cite{Varma:2023}. In a complementary direction, the complexity of simulating non-Hermitian dynamics for a collection of free $\mathcal{PT}$-symmetric systems via post-selection of an underlying interacting spin Hamiltonian was investigated in~\cite{Varma:2020qht}. This work highlights both the resource overhead and conceptual subtleties involved in embedding effective non-Hermitian evolution within fully Hermitian many-body dynamics.}} Construction of uncertainty relations for finite dimensional $\mathcal{PT}$-symmetric quantum systems was discussed in~\cite{Shukla:2022kda}. {A clear signature of a $\mathcal{PT}$-symmetry breaking phase transition has been identified in non-Hermitian lattice fermion models through eigenstate entanglement entropy, which exhibits an exponential divergence with system size at the transition point~\cite{Modak:2021hyn}.} 
 Temporal correlations across a  $\mathcal{PT}$ transition and quantum speed limit have been studied in~\cite{Varma:2022net}.
 {It has also been shown that classical polymeric unzipping transitions, including DNA unzipping, can be interpreted as generalized $\mathcal{PT}$-symmetry breaking transitions within an equivalent non-Hermitian quantum framework~\cite{Pal:2022gzg}.} {From a distinct topological perspective, it was recently shown that Hermitian topological phase transitions manifest as transitions between distinct knot topologies of complex eigenvalue spectra in associated non-Hermitian Hamiltonians, even in the absence of exceptional points~\cite{Hajong:2025hbh}.} 
Observation of $\mathcal{PT}$ transition and connection to inhomogeneous polarization transformation in non-Hermitian optical beam shifts has been reported in~\cite{Modak:2023qxt}.
For the $\mathcal{PT}$-symmetric Hamiltonian with the modified inner product, it was shown that the Hellmann-Feynmann theorem is valid in the $\mathcal{PT}$-broken phase, $\mathcal{PT}$-unbroken phase and at the exceptional point~\cite{Hajong:2024owo}. 
A Hermitian bypass to a non-Hermitian theory is proposed in~\cite{Bhasin:2023huf}.
Some consequences of non-Hermitian coupling in a spin-boson system have been discussed in~\cite{Das:2026}.
In optics~\cite{Zyablovsky_2014} and photonic context~\cite{Regensburger:2012knk} also, $\mathcal{PT}$-symmetric systems have been studied. Different forms of $\mathcal{P}$ and $\mathcal{T}$ operators are covered in~\cite{Rath:2020} and can be employed to check for $\mathcal{PT}$-symmetry of  a given Hamiltonian.

In the context of high energy physics, $\mathcal{PT}$-symmetric fermionic particle oscillations in even-dimensional representations have been studied in~\cite{Chen:2024bya}. Aspects of  relativistic non-Hermitian quantum mechanics have been discussed in~\cite{Jones-Smith:2009qeu,Jones:2010}. 
{{In~\cite{Mannheim:2015hto}, it was proposed that anti-linearity rather than Hermiticity should be a guiding principle for any quantum theory. Extension of the $\mathcal{CPT}$ theorem to non-Hermitian Hamiltonians and unstable states has been studied in~\cite{
Mannheim:2015tlu}.
In Ref.~\cite{Raval:2018kqg}, it was demonstrated that transition from deconfinement to confinement can be interpreted  as a $\mathcal{PT}$-phase transition. There are also studies on symmetry properties of non-Hermitian $\mathcal{PT}$-symmetric quantum field theories~\cite{Millington:2019dfn} as well as on discrete spacetime symmetries and particle mixing in non-Hermitian scalar quantum field theories~\cite{Alexandre:2020gah,Alexandre:2023afi}. In a recent work on $\mathcal{PT}$-symmetric quantum mechanics, ideas of time advance and probability conservation have been put forth~\cite{Mannheim:2025ztp}. In neutrino physics, non-Hermitian quantum dynamics provides a framework for describing both neutrino mixing and decay~\cite{Chattopadhyay:2021eba,Chattopadhyay:2022ftv,Dixit:2022izn,Parveen:2024cff}. This  opens up possibilities of distinguishing between Dirac and Majorana neutrinos~\cite{Dixit:2022izn,Soni:2023njf,Parveen:2024cff}. 
In particular, the treatment of  neutrino mixing with the inclusion of non-Hermitian terms is of current interest and has been investigated in~\cite{Ohlsson:2015xsa,Ohlsson:2019noy,Ohlsson:2020gxx,Ohlsson:2020idi,Alexandre:2023afi}. }}

Going beyond the Hermitian Hamiltonians, it may also be mentioned that there is another class of Hamiltonians referred to as the pseudo-Hermitian Hamiltonians. A Hamiltonian $\Tilde{\mathcal{H}}$ is pseudo-Hermitian if there exists an invertible Hermitian operator $\eta$ such that $\Tilde{\mathcal{H}}^\dagger = \eta \Tilde{\mathcal{H}} \eta^{-1}$~\cite{Mostafazadeh:2001jk, Mostafazadeh:2001nr,Mostafazadeh:2002id,Mostafazadeh:2006xuq,Mostafazadeh:2008pw}. This property ensures that pseudo-Hermitian systems can have real eigenvalue spectra under certain conditions, much like the $\mathcal{PT}$-symmetric systems in the unbroken regime. In fact, $\mathcal{PT}$-symmetric Hamiltonians are a subset of pseudo-Hermitian Hamiltonians.

In this work, we consider a general non-Hermitian Hamiltonian for two neutrino flavors. It may be noted that $\mathcal{PT}$-symmetry is respected for certain choice of parameters. For this case, we develop a mathematical framework using the $\mathcal{G}$ inner product~\cite{Ju:2019kso,Tzeng:2020mwv,Mannheim:2017apd,Shi:2009pc,Kleefeld:2009vd,Shukla:2022kda}. This {inner product} defined by positive-definite metric operator $\mathcal{G}$ is constructed from the bi-orthonormal set of left eigenvectors and right eigenvectors of the $\mathcal{PT}$-symmetric non-Hermitian Hamiltonian. With this approach, the probabilities in the $\mathcal{PT}$-unbroken regime as well as the $\mathcal{PT}$-broken regime do not necessarily lead to probability conservation. We then adopt another framework using  the density matrix prescription proposed by Brody and Graefe~\cite{Brody:2012nxf} (see also~\cite{Varma:2020tqu}). {\it{We show that this prescription  leads to consistent mathematical framework (probabilities being conserved) by imposing $\mathrm{Tr}(\rho)=1$. }}

 The plan of the article is as follows. In Sec.~\ref{sec:formalism}, we present the theoretical framework for two flavor neutrino oscillations in the presence of a non-Hermitian Hamiltonian and discuss the conditions under which the system exhibits $\mathcal{PT}$-symmetry. In Sec.~\ref{sec:G-metric-approach}, we develop the formalism based on the bi-orthonormal inner product defined by the positive-definite metric operator $\mathcal{G}$ and compute the oscillation probabilities in the regimes of interest ($\mathcal{PT}$-unbroken and $\mathcal{PT}$-broken). It turns out that the probabilities are not conserved in either regime.   In Sec.~\ref{sec:density-matrix-approach}, we introduce the density matrix prescription proposed by Brody and Graefe, which allows us to consistently describe the time evolution of the system while ensuring  probability conservation.  Finally, we show the graphical results in Sec.~\ref{sec:result}. We summarize in  Sec.~\ref{sec:summary}. In Appendix~\ref{procedure}, we write down the detailed procedure followed in the two approaches. In Appendix~\ref{numerator}, we provide an expression for the time evolved density matrix used for computing probabilities in Sec.~\ref{sec:density-matrix-approach}. Using conversion factors, we express the probability expressions (Sec.~\ref{sec:formalism}) in Appendices~\ref{Prob_G_factor} and \ref{Prob_density_matrix_factor}.
\section{Framework for  two flavor neutrino oscillations in presence of non-Hermitian dynamics}\label{sec:formalism}
In the ultra-relativistic limit, the two neutrino flavor states can be mapped to a two level quantum system with distinct energy eigenvalues, $E_i \approx p + m_i^2/2p$, 
along with the assumption of equal fixed momenta (or energy)~\cite{Mehta:2009ea}. 
The Dirac equation for two flavor neutrinos (antineutrinos)  can be reduced to a \sch like equation~\cite{Raffelt:1996wa} with the following Hamiltonian
\bea {\mathcal
 H} &=&  \dfrac{1}{4 E} \left(
\begin{array}{cc}
 \mbar -  \Dm \,\cos  2\theta  & \Dm \, \sin  2\theta \\
   \Dm \, \sin  2\theta  &    \mbar + \Dm \, \cos  2\theta
\end{array}
\right)\, . \label{nuham} \eea
where, $\mbar = m_i^2 + m_j^2$, $\Dm = m_i^2-m_j^2$ ($i,j$ refer to the label of the two energy eigenstates).
$\theta$ is the mixing angle in vacuum. The Hamiltonian in Eq.~\eqref{nuham} is Hermitian. The off-diagonal form of the (Hermitian) Hamiltonian in  flavor basis leads to flavor oscillations of neutrinos, which is the mechanism that mixes the neutrinos of different flavors while preserving the lepton number. The above (Hermitian) Hamiltonian is augmented by a non-Hermitian part, 
 \bea \mathcal{H} &=&  \dfrac{1}{4 E}\left[ \left(
\begin{array}{cc}
 \mbar -  \Dm \,\cos  2\theta  & \Dm \, \sin  2\theta \\
   \Dm \, \sin  2\theta  &    \mbar + \Dm \, \cos  2\theta
\end{array}
\right) + \begin{pmatrix}
\kappa e^{i\varphi} & \sigma e^{i\chi} \\
\sigma e^{-i\chi} & \kappa e^{-i\varphi}
\end{pmatrix}
\right] \, , \label{NH-H}
\eea
where, $\kappa, \sigma, \varphi$ and $\chi$ are real parameters. Note that the second term in Eq.~\eqref{NH-H} takes a symmetric form for $\chi = 0$ and respects  $\mathcal{PT}$-symmetry\footnote{Under the parity transformation $\mathcal{P}$, the  dynamical variables transform as $x \to -x$,  $p \to -p $, while time-reversal $\mathcal{T}$ is defined as $ x \to x$, $p \to -p$ and $i \to -i$. A Hamiltonian $\mathcal{H}$ is said to be $\mathcal{PT}$-symmetric if it commutes with the combined $\mathcal{PT}$ operation, \ie, if $[\mathcal{H}, \mathcal{PT}] = 0$. }.
The eigenvalues of the Hamiltonian are given by
\bea
E_{\pm} &=& \dfrac{1}{4 E}\left(  \kappa \cos \varphi +  \mbar \pm \sqrt{ {(\sigma + \Dm \sin 2 \theta)}^2 +  {(-i\kappa \sin \varphi + {\Dm \cos 2 \theta})}^2} \right) \, . \label{eigenvalue-general} 
\eea
Note that for the Hamiltonian defined in Eq.~\eqref{NH-H}, the eigenvalues are complex and the eigenvectors are no longer orthonormal under the conventional Dirac inner product. This calls for devising proper treatment such as using  a bi-orthonormal basis or the density matrix prescription by Brody and Graefe~\cite{Brody:2012nxf} (see also~\cite{Varma:2020tqu}). In Sec.~\ref{sec:G-metric-approach}, we define a new inner product defined by an appropriate metric operator $\mathcal{G}$. The Hamiltonian in Eq.~\eqref{NH-H} respects $\mathcal{PT}$-symmetry (for the special case when $\theta = \pi/4$ and $\chi = 0$ in Eq.~\eqref{NH-H}, we have $[\mathcal{H}, \mathcal{PT}] = 0$\footnote{In this case, the Hamiltonian is $\mathcal{PT}$-symmetric for two specific choices of the $\mathcal{P}$ and $\mathcal{T}$ operators: the first one is when the parity operator is  $\mathcal{P} = \sigma_x$ and the time-reversal operator is $\mathcal{T} = \mathcal{I}\mathcal{K}$, which is $\mathcal{T}$ even case, where $\mathcal{I}$ is the identity matrix and $\mathcal{K}$ is complex conjugation operator and the second choice is when $\mathcal{P} = \sigma_z$ and $\mathcal{T} = i \sigma_y\mathcal{K}$ which is $\mathcal{T}$ odd case ~\cite{Rath:2020,Jones-Smith:2009qeu}.}). The $\mathcal{PT}$-symmetry allows us to define the two distinct regime: the $\mathcal{PT}$-unbroken regime (real eigenvalues) and the $\mathcal{PT}$-broken regime (complex conjugate pairs of eigenvalues). The general non-Hermitian case is discussed in Sec.~\ref{sec:density-matrix-approach} where we have used the density matrix prescription proposed by Brody and Graefe~\cite{Brody:2012nxf} (see also~\cite{Varma:2020tqu}).
 \subsection{The $\mathcal{G}$ metric Approach} \label{sec:G-metric-approach}
A characteristic feature of non-Hermitian systems is that the eigenvectors of $\mathcal{H}$ are generally not orthonormal. The Hamiltonian $\mathcal{H}$ and its Hermitian conjugate $\mathcal{H}^\dagger$ possess distinct set of eigenvectors, denoted by $\ket{u_i}$ (right eigenvectors) and $\ket {v_i}$ (left eigenvectors) given by
\bea
\mathcal{H}\,\ket{u_i} &=& E_i\, \ket{u_i}\, , \qquad \mathcal{H}^\dagger \, \ket{v_i} = E_i^* \,\ket{v_i} \, .\label{left-right-eigenvectors}
\eea
These eigenvectors form a bi-orthonormal basis and obey
\bea
\scx{v_i}{u_j}  &=& \delta_{ij}\, , \qquad \sum_i \ket{u_i} \bra{v_i} = 1 \, . \label{orthonormality-completeness}
\eea\noindent
These right eigenvectors and left eigenvectors together form the complete bi-orthonormal basis. The left eigenvectors and right eigenvectors are related via the metric operator $\mathcal{G}$~\cite{Ju:2019kso,Tzeng:2020mwv,Mannheim:2017apd,Shi:2009pc,Kleefeld:2009vd,Shukla:2022kda}
\bea
\ket{v_i} &=& \mathcal{G} \,\ket{u_i} \, .
\eea
 $\mathcal{G}$ is given by
\bea
 \mathcal{G} &=& \sum_i \ket{v_i} \bra{v_i} = \left[ \sum_i \ket{u_i}\bra{u_i} \right]^{-1} \, ,\label{G}
\eea 
Using Eq.~\eqref{G}, the bi-orthonormality condition (Eq.~\eqref{orthonormality-completeness}) becomes
\bea
\scx{v_i}{u_j} &=& \me{\mathcal{G}}{u_i}{u_j} = \scx{u_i}{u_j}_{\mathcal{G}} = \delta_{ij}\, . \label{normalization}
\eea 
Thus, in order to describe the non-Hermitian systems, we define a generalized inner product referred to as the $\mathcal{G}$ inner product, given by
\bea
\scx{\psi}{\phi}_{\mathcal{G}} &=& \me{\mathcal{G}}{\psi}{\phi} \, . \label{G-inner-product}
\eea
where, $\ket{\psi}$ and $\ket{\phi}$ are any two state vectors. 
Let us now discuss the $\mathcal{PT}$-symmetric case {{\ie, when $\theta = \pi/4$ and $\chi =0$}}. In this case,  the eigenvalues are given by 
\bea
E_{\pm} &=& \dfrac{1}{4E}\left(\mbar + \kappa \cos \varphi \pm \sqrt{{(\sigma + \Dm)}^2 - {\kappa}^2 \sin^2 \varphi}\right) \, .\label{eigenvalues}
\eea \noindent

We can identify the following regimes of interest:
\bi
     \item \textbf{$\mathcal{PT}$-unbroken regime :}  When ${(\sigma + \Dm)}^2 > {\kappa}^2 {\sin}^2 \varphi$, the two  eigenvalues are real.
     
     \item \textbf{$\mathcal{PT}$-broken regime :}  When ${(\sigma + \Dm)}^2 < {\kappa}^2 {\sin}^2 \varphi$, the square root term becomes imaginary and the eigenvalues occur as complex conjugate pairs. 
     
\begin{figure}[t]
\centering
\includegraphics[width=3.5in]{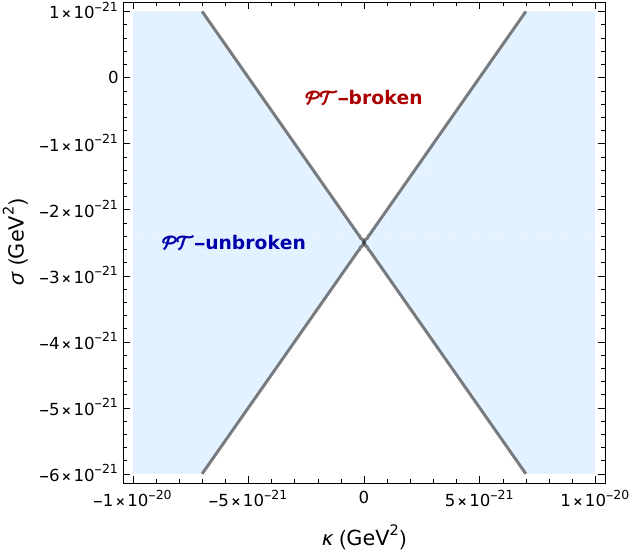}
\caption{\footnotesize{Regimes of interest  shown in $\kappa-\sigma$ plane for the $\mathcal{PT}$-symmetric case. 
The $\mathcal{PT}$-unbroken regime is shown as blue shaded region, while the $\mathcal{PT}$-broken regime is shown in white. The boundary represents the exceptional point. We have taken $ \Dm = 2.5 \times 10^{-3} \, \text{eV}^2, \varphi = \pi/6$.
}} 
\label{fig_phase}
\end{figure}

     \item \textbf{Exceptional point : }  When ${(\sigma + \Dm)}^2 = {\kappa}^2 {\sin}^2 \varphi$, the eigenvalues along with
the eigenvectors coalesce~\cite{Berry:2004ypy,Heiss:2012dx}. At this point, the Hamiltonian cannot be diagonalized, marking a  transition between $\mathcal{PT}$-unbroken regime and $\mathcal{PT}$-broken regime.
 \ei 
 {In Fig.~\ref{fig_phase}, we depict the regimes of interest in the $\kappa-\sigma$ plane. }
We shall now obtain the expression for probabilities.
We next describe the formalism for the $\mathcal{PT}$-unbroken regime.
\subsection{$\mathcal{PT}$-unbroken regime}\label{PT-unbroken} 
When ${(\sigma + \Dm)}^2 > \kappa^2 \sin^2 \varphi$, we are in the $\mathcal{PT}$-unbroken regime in which the eigenvalues are real. Here, the eigenstates of  $\mathcal{H}$ are simultaneous eigenstates of $\mathcal{PT}$~\cite{Bender:2007nj}. The normalized right eigenvectors are
\bea
\ket{u_+} &=& \dfrac{1}{\sqrt{2 \cos \tau}} \begin{pmatrix}
e^{ i \tau/2} \\
 e^{- i \tau/2}
\end{pmatrix} \quad \text{and} \quad \ket{u_-} = \dfrac{1}{\sqrt{2 \cos \tau}} \begin{pmatrix}
e^{- i \tau/2} \\
- e^{ i \tau/2}
\end{pmatrix} \, .\label{u-unbr}
\eea
The normalized left eigenvectors are
\bea
\ket{v_+} &=& \dfrac{1}{\sqrt{2 \cos \tau}} \begin{pmatrix}
e^{ -i \tau/2} \\
 e^{i \tau/2}
\end{pmatrix} \quad \text{and} \quad \ket{v_-} = \dfrac{1}{\sqrt{2 \cos \tau}} \begin{pmatrix}
e^{i \tau/2} \\
- e^{- i \tau/2}
\end{pmatrix} \, ,\label{v-unbr}
\eea
where, $\sin \tau = \kappa \sin \varphi/(\Dm + \sigma)$.
Using Eq.~\eqref{G} and Eq.~\eqref{u-unbr} (right eigenvectors), we can construct the $\mathcal{G}$
\bea
\mathcal{G} &=& \begin{pmatrix}
\sec \tau & -i \tan \tau \\
i \tan \tau & \sec \tau
\end{pmatrix} \, . \label{G-unbroken}
\eea
It may be noted that the normalized right eigenvectors $\ket{u_+}$ and $\ket{u_-}$ corresponds to the mass eigenstates. In order to discuss neutrino flavor oscillations we need to go from mass eigenstates to flavor states. Let us denote the flavor states by $\ket{u_a}$ and $\ket{u_b}$ which can be expressed in terms of mass eigenstates as follows
\bea
\ket{u_a} &=& {(\mathcal{A}^{-1})}_{a_+} \ket{u_+} + {(\mathcal{A}^{-1})}_{a_-} \ket{u_-} = \begin{pmatrix} 1 \\ 0 \end{pmatrix} \, , \nonumber\\
\ket{u_b} &=& {(\mathcal{A}^{-1})}_{b_+} \ket{u_+} + {(\mathcal{A}^{-1})}_{b_-} \ket{u_-} =  \begin{pmatrix} 0 \\ 1 \end{pmatrix} \, . \label{flavor1}
\eea
Here, $\mathcal{A}^{-1}$ is the similarity transformation matrix constructed from the right eigenvectors $\ket{u_+}$ and $\ket{u_-}$ of $\mathcal{H}$ and takes the following form
\bea
\mathcal{A}^{-1} &=& \dfrac{1}{\sqrt{2 \cos \tau}} \begin{pmatrix} 
e^{i \tau/2} & e^{-i \tau/2} \\
e^{-i \tau/2} & -e^{i \tau/2}
\end{pmatrix} \, . \label{A-unbr}
\eea
The time evolution of the flavor states is given by
\bea
\ket{u_a(t)} &=& {(\mathcal{A}^{-1})}_{a_+} \,e^{-i E_+ t}\, \ket{u_+} + {(\mathcal{A}^{-1})}_{a_-} \,e^{-i E_- t}\, \ket{u_-} = \dfrac{e^{-i \omega t}}{\cos \tau} \begin{pmatrix}
\cos (\tau - \zeta t)  \\
- i \sin \zeta t
\end{pmatrix}\, , \nonumber \\
\ket{u_b(t)} &=& {(\mathcal{A}^{-1})}_{b_+} \,e^{-i E_+ t}\, \ket{u_+} + {(\mathcal{A}^{-1})}_{b_-} \,e^{-i E_- t}\, \ket{u_-}
 = \dfrac{e^{-i \omega t}}{\cos \tau} \begin{pmatrix}
- i \sin \zeta t \\
\cos (\tau + \zeta t)
\end{pmatrix} \, ,  \label{ut-unbr}
\eea
where, $\omega  = (\rho \cos \varphi\ + \mbar)/4E$  and $\zeta = ( \sqrt{{(\sigma + \Dm)}^2  - \kappa^2 \sin^2 \varphi})/4E$.
Using these flavor states of $\mathcal{H}$, we can compute the transition probability (see Eq.~\eqref{G-inner-product} and Eq.~\eqref{G-unbroken})
\bea
P_{ab} &=& \dfrac{|\scx {u_b}{ u_a(t)}_\mathcal{G}|^2}{\scx{u_b}{u_b}_\mathcal{G} \,\scx{u_a(t)}{u_a(t)}_\mathcal{G}} \, .\label{prob1}
\eea
Noting that $\scx{u_a}{u_a}_\mathcal{G} = \scx{u_b}{u_b}_\mathcal{G} = \sec \tau$ and $\scx{u_a(t)}{u_a(t)}_\mathcal{G} =\scx{u_b(t)}{u_b(t)}_\mathcal{G} = \sec \tau$.
Thus we have
\bea
P_{aa} &=& \cos^2 \zeta t \, , \nonumber \\
P_{ab} &=& \sin^2(\tau - \zeta t) \, , \nonumber \\
P_{ba} &=& \sin^2(\tau + \zeta t)\, , \nonumber \\
P_{bb} &=& \cos^2 \zeta t \, . \label{Prob-unbr}
\eea
Eq.~\eqref{Prob-unbr} can be expressed in terms of baseline ($L$) and energy ($E$) (see Appendix~\ref{Prob_G_factor}). 
It may be noted that  $P_{aa} + P_{ab} \neq 1$ and $P_{ba} + P_{bb} \neq 1$ implying non-conservation of probabilities. We next describe the formalism of $\mathcal{PT}$-broken regime. 
\subsection{$\mathcal{PT}$-broken regime} \label{PT-broken}
When ${(\sigma + \Dm)}^2 < \kappa^2 \sin^2 \varphi$, we are in the $\mathcal{PT}$-broken regime in which the eigenvalues are complex conjugate pairs. Here, the eigenstates of  $\mathcal{H}$ are no longer simultaneous eigenstates of $\mathcal{PT}$. The normalized right eigenvectors are
 \bea
 \ket{ u' _+} &=& \dfrac{1}{\sqrt{2 \sinh \tau'}}  \begin{pmatrix}
 e^{\tau'/2} \\  -i e^{- \tau'/2}
\end{pmatrix} \quad \text{and}  \quad
\ket{u' _- } = \dfrac{1}{\sqrt{2 \sinh \tau'}}  \begin{pmatrix}
i e^{-\tau'/2} \\  e^{\tau'/2}
\end{pmatrix} \, .\label{u-br}
\eea 
The normalized left eigenvectors are
\bea
\ket{v'_+} &=& \dfrac{1}{\sqrt{2 \sinh \tau'}}  \begin{pmatrix}
e^{\tau'/2} \\
i e^{-\tau'/2}
\end{pmatrix} \quad \text{and} \quad \ket{v'_-} = \dfrac{1}{\sqrt{2 \sinh \tau'}}  \begin{pmatrix}
- i e^{-\tau'/2} \\
 e^{\tau'/2}
\end{pmatrix} \, ,\label{v-br}
\eea
 where, $\cosh \tau' = \kappa \sin \varphi / (\Dm + \sigma)$. The similarity transformation matrix $\mathcal{A}'^{-1}$ can be constructed using Eq.~\eqref{u-br} and has the following form
\bea
\mathcal{A}'^{-1} &=& \dfrac{1}{\sqrt{2 \sinh \tau'}} \begin{pmatrix}
e^{ \tau'/2} & i e^{-\tau'/2}  \\ -i e^{- \tau'/2} &     e^{ \tau'/2}  \end{pmatrix} \, . \label{A-br}
\eea
Following the steps laid down in Sec.~\ref{PT-unbroken}, we can obtain the flavor states using the similarity transformation matrix $\mathcal{A}'^{-1}$
\bea
\ket{u'_a} &=& {(\mathcal{A'}^{-1})}_{a_+} \ket{u'_+} + {(\mathcal{A'}^{-1})}_{a_-} \ket{u'_-} = \begin{pmatrix} 1 \\ 0 \end{pmatrix} \, , \nonumber\\
\ket{u'_b} &=& {(\mathcal{A'}^{-1})}_{b_+} \ket{u'_+} + {(\mathcal{A'}^{-1})}_{b_-} \ket{u'_-} =  \begin{pmatrix} 0 \\ 1 \end{pmatrix} \, . \label{flavor2}
\eea
 The time evolution of the flavor states is then given  by
\bea
 \ket{u'_a(t)} &=&{(\mathcal{A'}^{-1})}_{a_+} \,e^{-i E'_+ t}\, \ket{u'_+} + {(\mathcal{A'}^{-1})}_{a_-} \,e^{-i E'_- t}\, \ket{u'_-} = \dfrac{e^{-i\omega t}}{ \sinh \tau'} \begin{pmatrix}
\sinh (\tau'+\zeta' t)  \\  -  i \sinh \zeta' t
\end{pmatrix} \, \nonumber \\
\ket{u'_b(t)} &=& {(\mathcal{A'}^{-1})}_{b_+} \,e^{-i E'_+ t}\, \ket{u'_+} + {(\mathcal{A'}^{-1})}_{b_-} \,e^{-i E'_- t}\, \ket{u'_-} = \dfrac{e^{-i\omega t}}{ \sinh \tau'} \begin{pmatrix}
 -  i \sinh \zeta' t \\  \sinh (\tau'-\zeta' t)    
 \end{pmatrix} \, , \label{ut-br}
\eea
where, $\omega = ( \rho \cos \varphi\ + \mbar)/4E$  and $\zeta' = ( \sqrt{\kappa^2 \sin^2 \varphi - {(\Dm + \sigma)}^2})/4E$. In $\mathcal{PT}$-broken regime metric operator $\mathcal{G}$ can be time-independent or time-dependent.

Let us compute the probabilities with a time-dependent form of $\mathcal{G}$. The time-dependent $\mathcal{G}$ can be found by considering the time-dependence of the eigenvectors in Eq.~\eqref{G}
\bea
\mathcal{G}_{t} = \dfrac{1}{ \sinh \tau'} \begin{pmatrix}
 \cosh (\tau' - 2\zeta' t) & - i  \cosh  2\zeta' t \\   i  \cosh 2\zeta' t &     \cosh (\tau' + 2\zeta' t) \end{pmatrix}\, . \label{G-t}
\eea
Using the flavor states of $\mathcal{H}$, we can compute the transition probability (see Eq.~\eqref{prob1} and Eq.~\eqref{G-t}). Noting that $\scx{u_a'}{u_a'}_{\mathcal{G}_{t}} =\cosh (\tau' - 2 \zeta' t)/\sinh \tau', \scx{u_b'}{u_b'}_{\mathcal{G}_{t}} = \cosh (\tau' + 2 \zeta' t)/\sinh \tau'$ and $\scx{u'_a(t)}{u'_a(t)}_{\mathcal{G}_{t}} = \scx{u'_b(t)}{u'_b(t)}_{\mathcal{G}_{t}} = \coth \tau'$, we obtain
\bea
P'_{aa} &=&  \dfrac{\cosh^2 (\tau'-\zeta' t)}{\cosh \tau' \cosh (\tau'- 2\zeta' t)}\, , \nonumber \\
P'_{ab} &=&   \dfrac{\cosh^2 \zeta' t}{\cosh \tau' \cosh (\tau'+ 2\zeta' t)}\, , \nonumber\\
P'_{ba} &=&  \dfrac{\cosh^2  \zeta' t}{\cosh \tau' \cosh (\tau'- 2\zeta' t)}\, , \nonumber\\
P'_{bb} &=&  \dfrac{\cosh^2  (\tau' + \zeta' t) }{\cosh \tau' \cosh(\tau' + 2 \zeta' t) }\, .\label{Prob-br}
\eea
Eq.~\eqref{Prob-br} can be expressed in terms of $L$ and $E$ (see Appendix~\ref{Prob_G_factor}).
It may be noted that the probabilities are not conserved, \ie, $P'_{aa} + P'_{ab} \neq 1$ and $P'_{ba} + P'_{bb} \neq 1$. 
\subsection{Density Matrix Prescription by Brody and Graefe} \label{sec:density-matrix-approach}
Let us consider the time evolution of a general two level quantum system described by a density operator. Within this description, the neutrino system can be treated as an open quantum system that may interact with an external environment. At the production point, neutrinos are assumed to be in  pure flavor state. During propagation the system experience small perturbations that effectively encode interaction with environment. We assume that the dynamics is governed by a non-Hermitian Hamiltonian of the form $\mathcal{H} = \mathcal{B} - i \mathcal{C}$, where $\mathcal{B}$ and $\mathcal{C}$ are Hermitian matrices. The time evolution equation for the density matrix is given by~\cite{Brody:2012nxf} (see also~\cite{Varma:2020tqu})
\bea
\dfrac{d \rho}{d t} = - i [\mathcal{B},\,\rho] - \{\mathcal{C},\,\rho\} + 2 (\textbf{Tr} \rho\,\mathcal{C}) \rho   \, .\label{density-eq}
\eea
The first term $i [\mathcal{B},\,\rho]$, describes the unitary part of the evolution associated with the Hermitian matrix $\mathcal{B}$. The second term $\{\mathcal{C},\,\rho\}$, arises from the anti-Hermitian part of the Hamiltonian and introduces non-unitary contribution to the evolution which may be interpreted as effective gain or loss in the system. The last term ensures preservation of the normalization condition, $\textbf{Tr} (\rho) =1$. In particular, the anti-commutator term alone would modify the trace of the density matrix, while the last term compensates this contribution, keeping the trace constant during the evolution. Consequently, the dynamics remains trace preserving. The solution of Eq.~\eqref{density-eq} is given by
\bea
\rho_{k}(t) = \dfrac{e^{-i\mathcal{H}t} \rho_{k}(0)e^{i\mathcal{H}^{\dagger}t}}{\textbf{Tr}(e^{-i\mathcal{H}t} \rho_{k}(0)e^{i\mathcal{H}^{\dagger}t})} \, ,  \label{Density-t}
\eea
where, $\rho_{k}(0)$ corresponds to a pure state, $\rho_{k}(0) = \ket{\nu_{k}}\bra{\nu_{k}}$, $\ket{\nu_{k}}$ ($k = a,b$)  being the eigenvector of the Hermitian part of the Hamiltonian (Eq.~\eqref{nuham}). The above expression represents the normalized time evolution of the density matrix in the presence of a non-Hermitian Hamiltonian. The operator $e^{-i\mathcal{H}t} \rho_{k}(0)e^{i\mathcal{H}^{\dagger}t}$ describes the non-unitary evolution of the initial state $\rho_{k}(0)$. Since the Hamiltonian is not Hermitian, the evolution operator is not unitary and the trace of the density matrix is not automatically preserved. The normalization factor in the denominator, $\textbf{Tr}(e^{-i\mathcal{H}t} \rho_{k}(0)e^{i\mathcal{H}^{\dagger}t})$ ensures that the evolved density matrix satisfies the condition, $\textbf{Tr} (\rho(t)) =1$. From Eq.~\eqref{nuham},
\bea
\ket{\nu_a} &=&  \begin{pmatrix}
-\cos \theta \\
 \sin \theta
\end{pmatrix} \quad \text{and} \quad \ket{\nu_b} =  \begin{pmatrix}
\sin \theta \\
\cos \theta
\end{pmatrix} \, .\label{initial-state}
\eea
The density matrices corresponding to the initial states $\ket{\nu_a}$ and $\ket{\nu_b}$ are given by
\bea
\rho_{a}(0) = \begin{pmatrix}
    \cos^{2} \theta & - \sin \theta \cos \theta\\
    - \sin \theta \cos \theta & \sin^{2} \theta 
\end{pmatrix}
\quad \text{and} \quad \rho_{b}(0) =  \begin{pmatrix}
    \sin^{2} \theta &  \sin \theta \cos \theta\\
     \sin \theta \cos \theta & \cos^{2} \theta 
\end{pmatrix} \, , \label{pure-state}
\eea
The probability is given by
\bea
P_{ab} = \textbf{Tr}{(\rho_{b}(0)\rho_{a}(t))} \, ,
\label{Prob-density-approach-general}
\eea
where, $\rho_{a}(t)$ and $\rho_{b}(t)$ is calculated using Eq.~\eqref{Density-t} (see Appendix~\ref{numerator}). The probability expressions are (see Appendix~\ref{procedure})
\bea
P_{aa} &=& [(\cos 2 \gamma t - \cosh 2 \xi t )(\cos 4 \theta(2 + \cos 2 \beta - \cosh 2 \alpha)- 4 \cos \beta \sin 4 \theta \sinh \alpha) \nonumber \\
&& - \cos 2 \gamma t (2 - 3 \cos 2 \beta 
- \cosh 2 \alpha) + \cosh 2 \xi t (2 + \cos 2 \beta + 3 \cosh 2 \alpha)\nonumber \\
&& - 4 \cos 2 \theta (\sin 2 \gamma t \sin 2 \beta + \sinh 2 \xi t \sinh 2 \alpha) + 8 \sin 2 \theta(\sin 2 \gamma t \sin \beta \sinh \alpha \nonumber \\
&& - \sinh 2 \xi t \cos \beta \cosh \alpha)]/[4\{2 \cosh 2 \xi t \cosh^2 \alpha - 2 \cos 2 \gamma t \sin^2 \beta \nonumber \\ 
&& - \cos 2 \theta (\sin 2 \gamma t \sin 2 \beta + \sinh 2 \xi t \sinh 2 \alpha) + 2 \sin 2 \theta (\sin 2 \gamma t \sin \beta \sinh \alpha \nonumber \\
&& - \sinh 2 \xi t \cos \beta \cosh \alpha)\}] \, ,\nonumber \\
P_{ab} &=& [(\cosh 2 \xi t - \cos 2 \gamma t)(2 - \cos 2\beta + \cosh 2 \alpha + \cos4 \theta(2 + \cos 2 \beta - \cosh 2 \alpha)  \nonumber \\
&& - 4 \cos \beta \sin 4 \theta \sinh \alpha)]/[4\{2\cosh^2 \alpha \cosh 2 \xi t - 2 \cos 2 \gamma t \sin^2 \beta \nonumber \\
&& - \cos 2 \theta(\sin 2 \gamma t \sin 2 \beta + \sinh 2 \alpha \sinh 2 \xi t) \nonumber + 2 \sin 2\theta(\sin 2 \gamma t \sin \beta \sinh \alpha \nonumber \\ 
&& -  \sinh 2 \xi t \cos \beta \cosh \alpha)\}] \, , \nonumber\\ 
P_{ba} &=& [(\cosh 2 \xi t - \cos 2 \gamma t)(2 - \cos 2\beta + \cosh 2 \alpha + \cos4 \theta(2 + \cos 2 \beta - \cosh 2 \alpha) \nonumber \\
&& - 4 \cos \beta \sin 4 \theta \sinh \alpha)] /[4\{2\cosh^2 \alpha \cosh 2 \xi t - 2 \cos 2 \gamma t \sin^2 \beta \nonumber \\
&& + \cos 2 \theta(\sin 2 \gamma t \sin 2 \beta + \sinh 2 \alpha \sinh 2 \xi t) - 2 \sin 2\theta(\sin 2 \gamma t \sin \beta \sinh \alpha \nonumber \\ 
&& -  \sinh 2 \xi t \cos \beta \cosh \alpha)\}]\, , \nonumber \\
P_{bb} &=& [(\cos 2 \gamma t - \cosh 2 \xi t )(\cos 4 \theta(2 + \cos 2 \beta - \cosh 2 \alpha)- 4 \cos \beta \sin 4 \theta \sinh \alpha) \nonumber \\
&& - \cos 2 \gamma t (2 - 3 \cos 2 \beta - \cosh 2 \alpha) 
+ \cosh 2 \xi t (2 + \cos 2 \beta + 3 \cosh 2 \alpha)\nonumber \\
&& + 4 \cos 2 \theta (\sin 2 \gamma t \sin 2 \beta + \sinh 2 \xi t \sinh 2 \alpha) - 8 \sin 2 \theta(\sin2 \gamma t \sin \beta \sinh \alpha \nonumber \\ 
&& - \sinh 2 \xi t \cos \beta \cosh \alpha)]/[4\{2 \cosh 2 \xi t \cosh^2 \alpha - 2 \cos 2 \gamma t \sin^2 \beta \nonumber \\
&& + \cos 2 \theta (\sin 2 \gamma t \sin 2 \beta + \sinh 2 \xi t \sinh 2 \alpha) - 2 \sin 2 \theta (\sin 2 \gamma t \sin \beta \sinh \alpha \nonumber \\
&& - \sinh 2 \xi t \cos \beta \cosh \alpha)\}] \, ,
\label{Prob-density-approach}
\eea
 where, $\gamma = (\sigma + \Dm \sin 2 \theta) \cosh \alpha \cos \beta/4E$ and $\xi = (\sigma + \Dm \sin 2 \theta) \sinh \alpha \sin \beta /4E$. From the above expressions, we note that  $P_{aa}+P_{ab}=1$ and $P_{ba}+P_{bb}=1$.

The $\mathcal{PT}$-symmetric limit corresponds to setting $\theta = \pi/4$ and $\chi=0$ in the Hamiltonian given by Eq.~\eqref{NH-H}. From Eq.~\eqref{Prob-density-approach}, the probabilities reduce to
\bea
P_{aa} &=& [\cos 2 \gamma t (-2 + \cos 2 \beta + \cosh 2 \alpha)
+ \cosh 2 \xi t (2 + \cos 2\beta + \cosh2\alpha) \nonumber \\
&& + 4(\sin 2 \gamma t \sin \beta \sinh \alpha - \sinh 2 \xi t \cos \beta \cosh \alpha  )] /[4(\cosh 2 \xi t \cosh^{2} \alpha  -\cos 2 \gamma t \sin^{2} \beta   
\nonumber \\
&& + \sin 2 \gamma t \sin \beta \sinh \alpha - \sinh 2 \xi t \cos \beta \cosh \alpha)] \, ,\nonumber \\
P_{ab} &=& [(\cosh 2 \xi t - \cos 2 \gamma t)(\cosh 2 \alpha - \cos 2\beta)]\nonumber \\
&&/[4(\cosh 2 \xi t \cosh^{2} \alpha  -\cos 2 \gamma t \sin^{2} \beta +  \sin 2 \gamma t \sin \beta \sinh \alpha - \sinh 2 \xi t \cos \beta \cosh \alpha)] \, ,\nonumber \\
P_{ba} &=& [(\cosh 2 \xi t - \cos 2 \gamma t)(\cosh 2 \alpha - \cos 2\beta )]
\nonumber \\
&&/[4(\cosh 2 \xi t \cosh^{2} \alpha  -\cos 2 \gamma t \sin^{2} \beta -  \sin 2 \gamma t \sin \beta \sinh \alpha + \sinh 2 \xi t \cos \beta \cosh \alpha)] \, ,\nonumber \\
P_{bb} &=& [\cos 2 \gamma t (-2 + \cos 2 \beta + \cosh 2 \alpha)
+ \cosh 2 \xi t (2 + \cos 2\beta + \cosh2\alpha) \nonumber \\
&& - 4(\sin 2 \gamma t \sin \beta \sinh \alpha - \sinh 2 \xi t \cos \beta \cosh \alpha  )] /[4(\cosh 2 \xi t \cosh^{2} \alpha   -\cos 2 \gamma t \sin^{2} \beta \nonumber \\
&& -  \sin 2 \gamma t \sin \beta \sinh \alpha + \sinh 2 \xi t \cos \beta \cosh \alpha)] \, .  \label{Prob-density-approach-PT-limit}
\eea
Here, the prescription using density matrices by Brody and Graefe~\cite{Brody:2012nxf} (see also \cite{Varma:2020tqu}) involves trace preserving evolution of the density operator and this leads to probabilities being conserved even for the general non-Hermitian case. However, the probabilities take distinct limiting values. This is a characteristic of non-Markovian behavior~\cite{Breuer:2009pku}.
 \begin{figure}[t]
\centering
\includegraphics[width=7in]{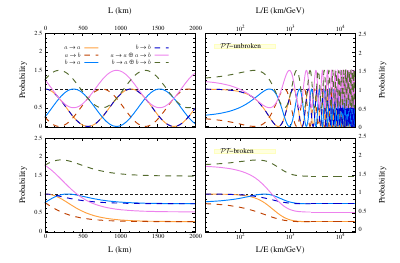}
\caption{\footnotesize{Left: Probability plotted as a function of $L$ for $E = 1 \,\text{GeV}$ for the $\mathcal{PT}$-unbroken case (top-left) and for the $\mathcal{PT}$-broken case (bottom-left). 
Right: Probability plotted as a function of $L/E$ \,$\text{km/GeV}$ for $\mathcal{PT}$-unbroken case (top-right) and for $\mathcal{PT}$-broken case (bottom-right). For the $\mathcal{PT}$-unbroken case, we have used $\sigma = 0$, $\tau = \pi/6$ and $\Dm = 2.5 \times 10^{-21}$ $\textrm{GeV}^2$. For the $\mathcal{PT}$-broken case, we have used $\sigma = 0$, $\tau' = \pi/6$ and $\Dm = 2.5 \times 10^{-21}$ $\textrm{GeV}^2$.
}} 
\label{fig_G}
\end{figure}
\section{Results and Discussion}
\label{sec:result}
In this section, we present  our results using the formalism given in Sec.~\ref{sec:formalism} (see Appendices~\ref{Prob_G_factor} and \ref{Prob_density_matrix_factor}) for probability expressions using appropriate conversion factors.

We first consider the case where probabilities are computed using $\mathcal{G}$ metric approach for the $\mathcal{PT}$-symmetric case (see Sec.~\ref{sec:G-metric-approach}). In Fig.~\ref{fig_G}, we plot the probability as a function of the baseline $L$ for fixed energy $E = 1$ GeV (left) and as a function of $L/E$ (right). The top panel  corresponds to the $\mathcal{PT}$-unbroken regime. If we take $\kappa$ and $\sigma$ to be $\mathcal{O}(10^{-21}\,\mathrm{GeV}^2)$, we are in the $\mathcal{PT}$-unbroken regime. The probabilities exhibit oscillatory behavior as a function of $L$ and $L/E$ (see Eq.~\eqref{Prob-unbr}). The bottom panel corresponds to the $\mathcal{PT}$-broken regime. If we take $\kappa$ to be $\mathcal{O}(10^{-20} \mathrm{GeV}^2)$ and $\sigma$  $\mathcal{O}(10^{-21} \mathrm{GeV}^2)$, we are in $\mathcal{PT}$-broken regime. For the $\mathcal{PT}$-symmetric case,  we conclude  that the probabilities are not conserved both in
$\mathcal{PT}$-unbroken regime and $\mathcal{PT}$-broken regime. 
\begin{figure}[t]
\centering
\includegraphics[width=7in]{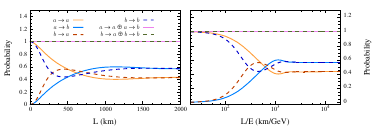}
\caption{\footnotesize{Left: Probability plotted as a function of $L$ for $E = 1 \,\text{GeV}$. Right: Probability plotted as a function of $L/E$. 
We have taken $\alpha = \pi/6, \beta= \pi/3, \theta = \pi/3, \Dm = 2.5 \times 10^{-3} \, \text{eV}^2, \sigma = 0$. 
}} 
\label{fig_density}
\end{figure}
\begin{figure}[t]
\centering
\includegraphics[width=7in]{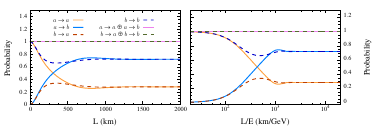}
\caption{\footnotesize{Left: Probability as function of $L$ for $E = 1 \,\text{GeV}$. Right: Probability as function of $L/E$. We have taken $\alpha = \pi/6, \beta= \pi/3, \Dm = 2.5 \times 10^{-3} \, \text{eV}^2, \sigma = 0$, $\theta = \pi/4$. 
}} 
\label{fig_compare}
\end{figure}
 
We next consider the case where probabilities are computed using  the density matrix prescription by Brody and Graefe~\cite{Brody:2012nxf} (see also \cite{Varma:2020tqu}) for the general non-Hermitian case (see Sec.~\ref{sec:density-matrix-approach}). In Fig.~\ref{fig_density}, we plot the probability as a function of the baseline $L$ for fixed value of energy $E = 1$ GeV (left) and as a function of $L/E$ (right). The density matrix prescription utilizes correct normalization factor which further leads to probabilities being conserved (see Eq.~\eqref{Prob-density-approach}) even for the general non-Hermitian case (see Eq.~\eqref{NH-H}).

In Fig.~\ref{fig_compare}, for the $\mathcal{PT}$- symmetric case, we show the probability as a function of baseline $L$ for fixed energy ($E = 1 \,\text{GeV}$) and $L/E$ using the density matrix prescription by Brody and Graefe~\cite{Brody:2012nxf} (see also \cite{Varma:2020tqu}). We notice that, in general, the  probabilities for different oscillation channels do not approach $1/2$ but there are two different limiting values. This is indicative of non-Markovian behavior~\cite{Breuer:2009pku}. 

\section{Conclusion}
\label{sec:summary}
In this work, we present a consistent framework to describe two flavor neutrino oscillations in presence of non-Hermitian dynamics. 
In existing literature,  $\mathcal{PT}$-symmetric Hamiltonians have been investigated~\cite{Ohlsson:2015xsa,Ohlsson:2019noy,Ohlsson:2020gxx,Ohlsson:2020idi} using the prescription of $\mathcal{CPT}$ inner product.
It was shown in~\cite{Ohlsson:2019noy,Ohlsson:2020idi}, that the framework using $\mathcal{CPT}$ {inner product}  works well  in the $\mathcal{PT}$-unbroken regime  leading  to consistent quantum theory for non-Hermitian systems. It may be noted that this claim relied on using the $\mathcal{CPT}$ eigenstates, which have no physical basis.

The key goal of the present work is to develop a framework which leads to probabilities being conserved for general non-Hermitian Hamiltonians.

In recent times, non-Hermitian dynamics in  neutrino oscillations  with decay has gained significant attention~\cite{Dixit:2022izn,Soni:2023njf,Parveen:2024cff,Chattopadhyay:2021eba,Chattopadhyay:2022ftv}. In these studies, the Hamiltonian is non-Hermitian but not necessarily $\mathcal{PT}$-symmetric. Moreover, a proper treatment (probabilities being semi-positive definite and conserved) of general non-Hermitian Hamiltonians in the context of neutrino oscillations is lacking. 

Here, we adopt two approaches to compute neutrino oscillation probabilities in presence of non-Hermitian dynamics, viz.,
\begin{itemize}
\item  Bi-orthonormal inner product defined by positive-definite metric operator $\mathcal{G}$~\cite{Ju:2019kso,Tzeng:2020mwv,Mannheim:2017apd,Shi:2009pc,Kleefeld:2009vd,Shukla:2022kda} and
\item   Density matrix prescription by Brody and Graefe~\cite{Brody:2012nxf} (see also \cite{Varma:2020tqu}).
\end{itemize}
In the 
  $\mathcal{G}$ metric approach (Sec.~\ref{sec:G-metric-approach}), the {inner product} is constructed from  left  and  right eigenvectors of the $\mathcal{PT}$-symmetric non-Hermitian Hamiltonian. 
Since the metric operator $\mathcal{G}$ is determined solely by the bi-orthonormal eigenstates of $\mathcal{H}$, this approach does not suffer from the operator choice ambiguity prevalent in the $\mathcal{CPT}$ {inner product} approach~\cite{Ohlsson:2020idi}. We would like to remark that in the $\mathcal{CPT}$ {inner product} approach, the construction of the inner product relies on the specific choice of the $\mathcal{P}$ and $\mathcal{T}$ operators. For a two-level $\mathcal{PT}$-symmetric system, multiple admissible choices of $\mathcal{P}$ and $\mathcal{T}$ exist and each choice leads to a different form of the $\mathcal{CPT}$ inner product~\cite{Rath:2020}. As a result, the probability expressions are dependent on the specific choice of $\mathcal{P}$ and $\mathcal{T}$ operators~\cite{Kleefeld:2009vd}.  However, it turns out that the $\mathcal{G}$ metric approach does not work well (probabilities are not conserved). We then consider an alternative framework~\cite{Brody:2012nxf} which is trace preserving by construction and ensures probability conservation even for the most general non-Hermitian case. In the density matrix prescription by Brody and Graefe~\cite{Brody:2012nxf} (see also \cite{Varma:2020tqu}), we start with an initial state density matrix corresponding to the Hermitian Hamiltonian and then evolve according to Eq.~\eqref{Density-t}. Using Eq.~\eqref{Prob-density-approach-general}, the probability is conserved and positive semi-definite (see Sec.~\ref{sec:density-matrix-approach}). {\it{We show that this prescription  leads to consistent mathematical framework (probabilities being conserved) by imposing $\mathrm{Tr}(\rho)=1$. }}

In Sec.~\ref{sec:result}, we present our results using the two approaches. We  find that  $\mathcal{G}$ metric approach does not works well (non-conservation of probability). Using the density matrix prescription by Brody and Graefe~\cite{Brody:2012nxf} (see also \cite{Varma:2020tqu}), we get conserved probabilities even for the general non-Hermitian case. Note that the limiting value of the probability  depends on the oscillation channel considered and is not equal to $1/2$. This  is indicative of non-Markovianity~\cite{Breuer:2009pku}.

The formalism developed here can be naturally extended to  the case of three neutrino flavors and include matter effects in neutrino propagation.

\section*{Acknowledgements}
 We  thank Sourin Das for suggesting the  density matrix approach by Brody and Graefe and for helpful discussions. The authors would like to acknowledge involvement of Sabila Parveen during the initial stages of this work. 
KR acknowledges financial support from Jawaharlal Nehru University. GH acknowledges the UGC-SRF for support. BPM acknowledges PDF grant of IOE, BHU for the year 2025-26. The research of PM is supported
 by the Inter-University Centre for Astronomy and Astrophysics (IUCAA), Pune and by the International Centre for Theoretical Sciences (ICTS), Bengaluru  through their
Associateship Programmes.  The research of PM has been partially supported by ANRF PAIR (ANRF/PAIR/2025/000029/PAIR-A). 

\appendix
\renewcommand{\theequation}{\thesection.\arabic{equation}}
\setcounter{equation}{0}
\counterwithin{figure}{section}
\numberwithin{equation}{section}
\renewcommand{\thesection}{\Alph{section}}
\renewcommand{\thesubsection}{\Alph{subsection}}
\section{Procedure for computing oscillation probabilities} \label{procedure}

Here, we summarize the procedure adopted to compute the oscillation probabilities using the two approaches considered in the present work.

\subsection*{Using the $\mathcal{G}$ metric approach}
Using the bi-orthonormal inner product defined by the positive-definite metric operator $\mathcal{G}$, we obtain the oscillation probabilities as follows:

\begin{itemize}

\item We write down the general non-Hermitian Hamiltonian $\mathcal{H}$ for two flavor neutrinos (see Eq.~\eqref{NH-H}). We consider $\mathcal{PT}$-symmetric form of the Hamiltonian  by substituting  $\theta = \pi/4$ and $\chi=0$ in Eq.~\eqref{NH-H}.

\item We compute the eigenvalues and eigenvectors (right and left eigenvectors of $\mathcal{H}$) (see Eq.~\eqref{left-right-eigenvectors}).

\item We construct the positive-definite metric operator $\mathcal{G}$ from the bi-orthonormal set of eigenvectors for the $\mathcal{PT}$-unbroken regime and $\mathcal{PT}$-broken regime (see Eq.~\eqref{G-unbroken} and Eq.~\eqref{G-t}).

\item We normalize the eigenvectors using the $\mathcal{G}$ inner product (see Eq.~\eqref{normalization}).

\item We write down the time evolved flavor states (see Eq.~\eqref{ut-unbr} and Eq.~\eqref{ut-br}) using $\mathcal{A}^{-1}$ (see Eq.~\eqref{A-unbr} and Eq.~\eqref{A-br}).

\item We finally compute the oscillation probabilities using Eq.~\eqref{prob1} for the $\mathcal{PT}$-unbroken regime and $\mathcal{PT}$-broken regime (see Eq.~\eqref{Prob-unbr} and Eq.~\eqref{Prob-br}).

\end{itemize}

\subsection*{Using the density matrix prescription by Brody and Graefe}
Using the the density matrix prescription proposed by Brody and Graefe~\cite{Brody:2012nxf}, we obtain the oscillation probabilities as follows:

\begin{itemize}

\item We write down the general non-Hermitian Hamiltonian $\mathcal{H}$ for two flavor neutrinos (see Eq.~\eqref{NH-H}). Note that the Hamiltonian is in general not $\mathcal{PT}$-symmetric.

\item We write down the time evolved density matrix using Eq.~\eqref{Density-t} (see Appendix~\ref{numerator}). Note that the initial flavor density matrix is given by Eq.~\eqref{pure-state}. 

\item We compute the oscillation probabilities using Eq.~\eqref{Prob-density-approach-general}.

\end{itemize}
\section{Numerator of time evolved density matrix} \label{numerator}
The expressions for the numerator of the time evolved density matrix $\rho_{a}(t)$ and $\rho_{b}(t)$ used in Eq.~\eqref{Density-t} are given by
\bea 
e^{-i\mathcal{H}t} \rho_{a}(0)e^{i\mathcal{H}^{\dagger}t} &=& e^{-i \omega t }\begin{pmatrix}
   i \tanh z \sin \Gamma t + \cos \Gamma t& - i \sin \Gamma t \sech z \\
   - i \sin \Gamma t \sech z & -i \tanh z \sin\Gamma t + \cos \Gamma t
\end{pmatrix}\nonumber \\
&& \times \begin{pmatrix}
    \cos^{2} \theta & - \sin \theta \cos \theta\\
    - \sin \theta \cos \theta & \sin^{2} \theta 
\end{pmatrix} \nonumber \\
&& \times e^{i \omega t }\begin{pmatrix}
   -i \tanh z^{*} \sin \Gamma^{*} t + \cos \Gamma^{*} t &  i \sin \Gamma^{*} t \sech z^{*} \\
    i \sin \Gamma^{*} t \sech z^{*} & i \tanh z^{*} \sin\Gamma^{*} t + \cos \Gamma^{*} t
\end{pmatrix} \, , \nonumber
\eea
\bea 
e^{-i\mathcal{H}t} \rho_{b}(0)e^{i\mathcal{H}^{\dagger}t}&=& e^{-i \omega t }\begin{pmatrix}
   i \tanh z \sin \Gamma t + \cos \Gamma t & - i \sin \Gamma t \sech z \\
   - i \sin \Gamma t \sech z & -i \tanh z \sin\Gamma t + \cos \Gamma t
\end{pmatrix} \nonumber \\
&& \times \begin{pmatrix}
    \sin^{2} \theta &  \sin \theta \cos \theta\\
     \sin \theta \cos \theta & \cos^{2} \theta 
\end{pmatrix} \nonumber \\
&& \times e^{i \omega t }\begin{pmatrix}
   -i \tanh z^{*} \sin \Gamma^{*} t + \cos \Gamma^{*} t &  i \sin \Gamma^{*} t \sech z^{*} \\
    i \sin \Gamma^{*} t \sech z^{*} & i \tanh z^{*} \sin\Gamma^{*} t + \cos \Gamma^{*} t
\end{pmatrix} \, . \nonumber
\eea
where, $z = \alpha + i \beta, z^{*} = \alpha - i \beta$, $\Gamma = (\sqrt{ {(\sigma + \Dm \sin 2 \theta)}^2 +  {(-i\kappa \sin \varphi + {\Dm \cos 2 \theta})}^2})/4E$. Since $\Gamma$ is complex, we can rewrite $\Gamma = \gamma + i \xi$, where $\gamma = (\sigma + \Dm \sin 2 \theta) \cosh \alpha \cos \beta/4E$ and $\xi = (\sigma + \Dm \sin 2 \theta) \sinh \alpha \sin \beta /4E$. $ \Gamma^{*} = \gamma - i \xi$. Here
\bea
\tanh z &=& (-i\kappa \sin \varphi +  \Dm \cos 2 \theta)/(\sqrt{{(\sigma + \Dm \sin 2\theta)}^{2} + {(-i\kappa \sin \varphi + \Dm \cos 2 \theta)}^{2}}) \, , \nonumber \\ 
\sech z &=& (\sigma  + \Dm \sin 2 \theta)/(\sqrt{{(\sigma + \Dm \sin 2\theta)}^{2} + {(-i \kappa \sin \varphi +  \Dm \cos 2 \theta)}^{2}}) \, , \nonumber \\
\tanh z^{*} &=& (i\kappa \sin \varphi + \Dm \cos 2 \theta)/({\sqrt{{(\sigma + \Dm \sin 2\theta)}^{2} + {(i\kappa \sin \varphi + \Dm \cos 2 \theta)}^{2}}}) \, , \nonumber \\ 
\sech z^{*} &=& (\sigma  + \Dm \sin 2 \theta)/({\sqrt{{(\sigma + \Dm \sin 2\theta)}^{2} + {(i\kappa\sin \varphi + \Dm \cos 2 \theta)}^{2}}})\, . \nonumber
\eea
\section{Probabilities using the $\mathcal{G}$ metric approach}
\label{Prob_G_factor}
Using appropriate conversion factors, we can express oscillation probabilities in the $\mathcal{PT}$-unbroken regime (Eq.~\eqref{Prob-unbr}) as
\bea
P_{aa} &=& \cos^2 \left(1.27 \cos \tau \dfrac{(\Dm + \sigma)}{\mathrm{eV^2}}\dfrac{L}{E} \dfrac{\textrm{GeV}}{\textrm{km}}\right)\, , \nonumber \\
P_{ab} &=& \sin^2\left(\tau - 1.27 \cos \tau \dfrac{(\Dm + \sigma)}{\mathrm{eV^2}}\dfrac{L}{E} \dfrac{\textrm{GeV}}{\textrm{km}}\right) \, , \nonumber \\
P_{ba} &=& \sin^2\left(\tau + 1.27 \cos \tau \dfrac{(\Dm + \sigma)}{\mathrm{eV^2}}\dfrac{L}{E} \dfrac{\textrm{GeV}}{\textrm{km}}\right)\, , \nonumber \\
P_{bb} &=& \cos^2 \left(1.27 \cos \tau \dfrac{(\Dm + \sigma)}{\mathrm{eV^2}}\dfrac{L}{E} \dfrac{\textrm{GeV}}{\textrm{km}}\right)\, . \label{prob-unbr-units}
\eea
We can express oscillation probabilities in the $\mathcal{PT}$-broken regime (Eq.~\eqref{Prob-br}) as
\bea
P'_{aa} &=&  \dfrac{\cosh^2 \left( \tau'- 1.27 \sinh \tau' \dfrac{(\Dm + \sigma)}{\mathrm{eV^2}}\dfrac{L}{E} \dfrac{\textrm{GeV}}{\textrm{km}}\right)}{\cosh \tau' \cosh \left(\tau'-  2.54 \sinh \tau' \dfrac{(\Dm + \sigma)}{\mathrm{eV^2}}\dfrac{L}{E} \dfrac{\textrm{GeV}}{\textrm{km}}\right)}\, , \nonumber\\
P'_{ab} &=&   \dfrac{\cosh^2  \left( 1.27 \sinh \tau' \dfrac{(\Dm + \sigma)}{\mathrm{eV^2}}\dfrac{L}{E} \dfrac{\textrm{GeV}}{\textrm{km}}\right)}{ \cosh \tau' \cosh \left(\tau'+  2.54 \sinh \tau' \dfrac{(\Dm + \sigma)}{\mathrm{eV^2}}\dfrac{L}{E} \dfrac{\textrm{GeV}}{\textrm{km}}\right)}\, , \nonumber \\
P'_{ba} &=& \dfrac{\cosh^2 \left( 1.27 \sinh \tau' \dfrac{(\Dm + \sigma)}{\mathrm{eV^2}}\dfrac{L}{E} \dfrac{\textrm{GeV}}{\textrm{km}}\right)}{ \cosh \tau' \cosh \left(\tau'-  2.54 \sinh \tau' \dfrac{(\Dm + \sigma)}{\mathrm{eV^2}}\dfrac{L}{E} \dfrac{\textrm{GeV}}{\textrm{km}}\right)} \, ,\nonumber\\
P'_{bb} &=&  \dfrac{\cosh^2 \left( \tau'+ 1.27 \sinh \tau' \dfrac{(\Dm + \sigma)}{\mathrm{eV^2}}\dfrac{L}{E} \dfrac{\textrm{GeV}}{\textrm{km}}\right)}{\cosh \tau' \cosh \left(\tau' +  2.54 \sinh \tau' \dfrac{(\Dm + \sigma)}{\mathrm{eV^2}}\dfrac{L}{E} \dfrac{\textrm{GeV}}{\textrm{km}}\right)} \, . \label{prob-br-H-units}
\eea
 \section{Probabilities using the density matrix prescription by Brody and Graefe}
\label{Prob_density_matrix_factor}
Using appropriate conversion factors, we can express oscillation probabilities (Eq.~\eqref{Prob-density-approach}) as
\bea
 P_{aa} &=& \Big[\Big(\cos \left(2.54 \cosh \alpha \cos \beta \dfrac{(\sigma + \Dm\sin^2 2 \theta)}{\mathrm{eV^2}}\dfrac{L}{E} \dfrac{\textrm{GeV}}{\textrm{km}}\right) \nonumber \\
 &&- \cosh  \left(2.54 \sinh \alpha \sin \beta \dfrac{(\sigma + \Dm\sin^2 2 \theta)}{\mathrm{eV^2}}\dfrac{L}{E} \dfrac{\textrm{GeV}}{\textrm{km}}\right)\Big(\cos 4 \theta\Big(2 + \cos 2 \beta - \cosh 2 \alpha\Big)
 \nonumber \\
 && - 4 \cos \beta \sin 4 \theta \sinh \alpha\Big)  - \cos \left(2.54 \cosh \alpha \cos \beta \dfrac{(\sigma + \Dm\sin^2 2 \theta)}{\mathrm{eV^2}}\dfrac{L}{E} \dfrac{\textrm{GeV}}{\textrm{km}}\right) \nonumber \\
 &&\Big(2 - 3 \cos 2 \beta 
 - \cosh 2 \alpha\Big)  + \cosh \left(2.54 \sinh \alpha \sin \beta \dfrac{(\sigma + \Dm\sin^2 2 \theta)}{\mathrm{eV^2}}\dfrac{L}{E} \dfrac{\textrm{GeV}}{\textrm{km}}\right)\nonumber \\
 && \Big(2 + \cos 2 \beta + 3 \cosh 2 \alpha\Big)- 4 \cos 2 \theta (\sin \left(2.54 \cosh \alpha \cos \beta \dfrac{(\sigma + \Dm\sin^2 2 \theta)}{\mathrm{eV^2}}\dfrac{L}{E} \dfrac{\textrm{GeV}}{\textrm{km}}\right)\nonumber \\
 && \sin 2 \beta  + \sinh \left(2.54 \sinh \alpha \sin \beta \dfrac{(\sigma + \Dm\sin^2 2 \theta)}{\mathrm{eV^2}}\dfrac{L}{E} \dfrac{\textrm{GeV}}{\textrm{km}}\right) \sinh 2 \alpha\Big) + 8 \sin 2 \theta \nonumber \\
 &&\Big(\sin \left(2.54 \cosh \alpha \cos \beta \dfrac{(\sigma + \Dm\sin^2 2 \theta)}{\mathrm{eV^2}}\dfrac{L}{E} \dfrac{\textrm{GeV}}{\textrm{km}}\right) \sin \beta \sinh \alpha \nonumber \\
&& - \sinh \left(2.54 \sinh \alpha \sin \beta \dfrac{(\sigma + \Dm\sin^2 2 \theta)}{\mathrm{eV^2}}\dfrac{L}{E} \dfrac{\textrm{GeV}}{\textrm{km}}\right) \cos \beta \cosh \alpha\Big)\Big]\nonumber \\
&&\Big/\Big[4\Big(2 \cosh \left(2.54 \sinh \alpha \sin \beta \dfrac{(\sigma + \Dm\sin^2 2 \theta)}{\mathrm{eV^2}}\dfrac{L}{E} \dfrac{\textrm{GeV}}{\textrm{km}}\right)\cosh^2 \alpha \nonumber \\
&&- 2 \cos \left(2.54 \cosh \alpha \cos \beta \dfrac{(\sigma + \Dm\sin^2 2 \theta)}{\mathrm{eV^2}}\dfrac{L}{E} \dfrac{\textrm{GeV}}{\textrm{km}}\right) \sin^2 \beta \nonumber \\ 
&& - \cos 2 \theta \Big(\sin \left(2.54 \cosh \alpha \cos \beta \dfrac{(\sigma + \Dm\sin^2 2 \theta)}{\mathrm{eV^2}}\dfrac{L}{E} \dfrac{\textrm{GeV}}{\textrm{km}}\right) \sin 2 \beta \nonumber \\
&& + \sinh \left(2.54 \sinh \alpha \sin \beta \dfrac{(\sigma + \Dm\sin^2 2 \theta)}{\mathrm{eV^2}}\dfrac{L}{E} \dfrac{\textrm{GeV}}{\textrm{km}}\right) \sinh 2 \alpha\Big) \nonumber \\ 
&& + 2 \sin 2 \theta \Big(\sin \left(2.54 \sinh \alpha \sin \beta \dfrac{(\sigma + \Dm\sin^2 2 \theta)}{\mathrm{eV^2}}\dfrac{L}{E} \dfrac{\textrm{GeV}}{\textrm{km}}\right) \sin \beta \sinh \alpha \nonumber \\
&& - \sinh \left(2.54 \sinh \alpha \sin \beta \dfrac{(\sigma + \Dm \sin^2 2 \theta)}{\mathrm{eV^2}}\dfrac{L}{E} \dfrac{\textrm{GeV}}{\textrm{km}}\right) \cos \beta \cosh \alpha \Big)\Big)\Big] \, ,\nonumber \eea
 \bea
 P_{ab} &=& \Big[\Big(\cosh \left(2.54 \sinh \alpha \sin \beta \dfrac{(\sigma + \Dm\sin^2 2 \theta)}{\mathrm{eV^2}}\dfrac{L}{E} \dfrac{\textrm{GeV}}{\textrm{km}}\right) \nonumber \\
 && - \cos \left(2.54 \cosh \alpha \cos \beta \dfrac{(\sigma + \Dm\sin^2 2 \theta)}{\mathrm{eV^2}}\dfrac{L}{E} \dfrac{\textrm{GeV}}{\textrm{km}}\right)\Big)\Big(2 - \cos 2\beta + \cosh 2 \alpha 
 \nonumber \\
 && + \cos4 \theta\Big(2 + \cos 2 \beta - \cosh 2 \alpha\Big)  - 4 \cos \beta \sin 4 \theta \sinh \alpha\Big)\Big] \nonumber \\
 &&\Big/\Big[4\Big(2\cosh^2 \alpha \cosh \left(2.54 \sinh \alpha \sin \beta \dfrac{(\sigma + \Dm\sin^2 2 \theta)}{\mathrm{eV^2}}\dfrac{L}{E} \dfrac{\textrm{GeV}}{\textrm{km}}\right) \nonumber \\
 && - 2 \cos \left(2.54 \cosh \alpha \cos \beta \dfrac{(\sigma + \Dm\sin^2 2 \theta)}{\mathrm{eV^2}}\dfrac{L}{E} \dfrac{\textrm{GeV}}{\textrm{km}}\right) \sin^2 \beta \nonumber \\
 && - \cos 2 \theta\Big(\sin  \left(2.54 \cosh \alpha \cos \beta \dfrac{(\sigma + \Dm\sin^2 2 \theta)}{\mathrm{eV^2}}\dfrac{L}{E} \dfrac{\textrm{GeV}}{\textrm{km}}\right) \sin 2 \beta \nonumber \\
 && + \sinh 2 \alpha \sinh \left(2.54 \sinh \alpha \sin \beta \dfrac{(\sigma + \Dm\sin^2 2 \theta)}{\mathrm{eV^2}}\dfrac{L}{E} \dfrac{\textrm{GeV}}{\textrm{km}}\right)\Big) \nonumber \\
 && + 2 \sin 2\theta\Big(\sin \left(2.54 \cosh \alpha \cos \beta \dfrac{(\sigma + \Dm\sin^2 2 \theta)}{\mathrm{eV^2}}\dfrac{L}{E} \dfrac{\textrm{GeV}}{\textrm{km}}\right) \sin \beta \sinh \alpha \nonumber \\ 
 && -  \sinh \left(2.54 \sinh \alpha \sin \beta \dfrac{(\sigma + \Dm\sin^2 2 \theta)}{\mathrm{eV^2}}\dfrac{L}{E} \dfrac{\textrm{GeV}}{\textrm{km}}\right) \cos \beta \cosh \alpha\Big)\Big)\Big] \, , \nonumber\eea
 \bea
 P_{ba} &=& \Big[\Big(\cosh \left(2.54 \sinh \alpha \sin \beta \dfrac{(\sigma + \Dm\sin^2 2 \theta)}{\mathrm{eV^2}}\dfrac{L}{E} \dfrac{\textrm{GeV}}{\textrm{km}}\right) \nonumber \\
&& - \cos \left(2.54 \cosh \alpha \cos \beta \dfrac{(\sigma + \Dm\sin^2 2 \theta)}{\mathrm{eV^2}}\dfrac{L}{E} \dfrac{\textrm{GeV}}{\textrm{km}}\right)\Big)\Big(2 - \cos 2\beta + \cosh 2 \alpha \nonumber \\
&& + \cos4 \theta\Big(2 + \cos 2 \beta - \cosh 2 \alpha\Big)  - 4 \cos \beta \sin 4 \theta \sinh \alpha\Big)\Big] \nonumber \\
&&\Big/\Big[4\Big(2\cosh^2 \alpha \cosh \left(2.54 \sinh \alpha \sin \beta \dfrac{(\sigma + \Dm\sin^2 2 \theta)}{\mathrm{eV^2}}\dfrac{L}{E} \dfrac{\textrm{GeV}}{\textrm{km}}\right) \nonumber \\
 && - 2 \cos \left(2.54 \cosh \alpha \cos \beta \dfrac{(\sigma + \Dm\sin^2 2 \theta)}{\mathrm{eV^2}}\dfrac{L}{E} \dfrac{\textrm{GeV}}{\textrm{km}}\right) \sin^2 \beta \nonumber \\
 && + \cos 2 \theta \Big(\sin  \left(2.54 \cosh \alpha \cos \beta \dfrac{(\sigma + \Dm\sin^2 2 \theta)}{\mathrm{eV^2}}\dfrac{L}{E} \dfrac{\textrm{GeV}}{\textrm{km}}\right) \sin 2 \beta \nonumber \\
 && + \sinh 2 \alpha \sinh \left(2.54 \sinh \alpha \sin \beta \dfrac{(\sigma + \Dm\sin^2 2 \theta)}{\mathrm{eV^2}}\dfrac{L}{E} \dfrac{\textrm{GeV}}{\textrm{km}}\right)\Big) \nonumber \nonumber \\
 && - 2 \sin 2\theta\Big(\sin \left(2.54 \cosh \alpha \cos \beta \dfrac{(\sigma + \Dm\sin^2 2 \theta)}{\mathrm{eV^2}}\dfrac{L}{E} \dfrac{\textrm{GeV}}{\textrm{km}}\right) \sin \beta \sinh \alpha \nonumber \\ 
 && -  \sinh \left(2.54 \sinh \alpha \sin \beta \dfrac{(\sigma + \Dm\sin^2 2 \theta)}{\mathrm{eV^2}}\dfrac{L}{E} \dfrac{\textrm{GeV}}{\textrm{km}}\right) \cos \beta \cosh \alpha\Big)\Big)\Big] \, , \nonumber\eea
 \bea
 P_{bb} &=& \Big[\Big(\cos \left(2.54 \cosh \alpha \cos \beta \dfrac{(\sigma + \Dm\sin^2 2 \theta)}{\mathrm{eV^2}}\dfrac{L}{E} \dfrac{\textrm{GeV}}{\textrm{km}}\right) \nonumber \\
 &&- \cosh  \left(2.54 \sinh \alpha \sin \beta \dfrac{(\sigma + \Dm\sin^2 2 \theta)}{\mathrm{eV^2}}\dfrac{L}{E} \dfrac{\textrm{GeV}}{\textrm{km}}\right)\Big(\cos 4 \theta\Big(2 + \cos 2 \beta - \cosh 2 \alpha \Big)
 \nonumber \\
 && - 4 \cos \beta \sin 4 \theta \sinh \alpha \Big)  - \cos \left(2.54 \cosh \alpha \cos \beta \dfrac{(\sigma + \Dm\sin^2 2 \theta)}{\mathrm{eV^2}}\dfrac{L}{E} \dfrac{\textrm{GeV}}{\textrm{km}}\right) \nonumber \\
 &&\Big(2 - 3 \cos 2 \beta - \cosh 2 \alpha \Big)  + \cosh \left(2.54 \sinh \alpha \sin \beta \dfrac{(\sigma + \Dm\sin^2 2 \theta)}{\mathrm{eV^2}}\dfrac{L}{E} \dfrac{\textrm{GeV}}{\textrm{km}}\right)\nonumber \\
 && \Big(2 + \cos 2 \beta + 3 \cosh 2 \alpha\Big)+ 4 \cos 2 \theta (\sin \left(2.54 \cosh \alpha \cos \beta \dfrac{(\sigma + \Dm\sin^2 2 \theta)}{\mathrm{eV^2}}\dfrac{L}{E} \dfrac{\textrm{GeV}}{\textrm{km}}\right)\nonumber \\
 && \sin 2 \beta  + \sinh \left(2.54 \sinh \alpha \sin \beta \dfrac{(\sigma + \Dm\sin^2 2 \theta)}{\mathrm{eV^2}}\dfrac{L}{E} \dfrac{\textrm{GeV}}{\textrm{km}}\right) \sinh 2 \alpha\Big) - 8 \sin 2 \theta \nonumber \\
 &&\Big(\sin \left(2.54 \cosh \alpha \cos \beta \dfrac{(\sigma + \Dm\sin^2 2 \theta)}{\mathrm{eV^2}}\dfrac{L}{E} \dfrac{\textrm{GeV}}{\textrm{km}}\right) \sin \beta \sinh \alpha \nonumber \\
&& - \sinh \left(2.54 \sinh \alpha \sin \beta \dfrac{(\sigma + \Dm\sin^2 2 \theta)}{\mathrm{eV^2}}\dfrac{L}{E} \dfrac{\textrm{GeV}}{\textrm{km}}\right) \cos \beta \cosh \alpha \Big)\Big]\nonumber \\
&&\Big/\Big[4\Big(2 \cosh \left(2.54 \sinh \alpha \sin \beta \dfrac{(\sigma + \Dm\sin^2 2 \theta)}{\mathrm{eV^2}}\dfrac{L}{E} \dfrac{\textrm{GeV}}{\textrm{km}}\right)\cosh^2 \alpha \nonumber \\
&&- 2 \cos \left(2.54 \cosh \alpha \cos \beta \dfrac{(\sigma + \Dm\sin^2 2 \theta)}{\mathrm{eV^2}}\dfrac{L}{E} \dfrac{\textrm{GeV}}{\textrm{km}}\right) \sin^2 \beta \nonumber \\ 
 && + \cos 2 \theta \Big(\sin \left(2.54 \cosh \alpha \cos \beta \dfrac{(\sigma + \Dm\sin^2 2 \theta)}{\mathrm{eV^2}}\dfrac{L}{E} \dfrac{\textrm{GeV}}{\textrm{km}}\right) \sin 2 \beta \nonumber \\
 && + \sinh \left(2.54 \sinh \alpha \sin \beta \dfrac{(\sigma + \Dm\sin^2 2 \theta)}{\mathrm{eV^2}}\dfrac{L}{E} \dfrac{\textrm{GeV}}{\textrm{km}}\right) \sinh 2 \alpha \Big) \nonumber \\ 
 && - 2 \sin 2 \theta \Big(\sin \left(2.54 \sinh \alpha \sin \beta \dfrac{(\sigma + \Dm\sin^2 2 \theta)}{\mathrm{eV^2}}\dfrac{L}{E} \dfrac{\textrm{GeV}}{\textrm{km}}\right) \sin \beta \sinh \alpha \nonumber \\
 && - \sinh \left(2.54 \sinh \alpha \sin \beta \dfrac{(\sigma + \Dm \sin^2 2 \theta)}{\mathrm{eV^2}}\dfrac{L}{E} \dfrac{\textrm{GeV}}{\textrm{km}}\right) \cos \beta \cosh \alpha\Big)\Big)\Big] \, .  
 \label{Prob-density-approach-units}
\eea

\bibliographystyle{JHEP}
\bibliography{ref}
\end{document}